\documentclass[aps,prd,preprint,groupedaddress,preprintnumbers,longbibliography]{revtex4-1}

\newcommand{\nn}{\nonumber}
\newcommand{\lb}{\left\lbrace}
\newcommand{\rb}{\right\rbrace}
\newcommand{\SL}{S_\Lambda}
\newcommand{\SIL}{S_{I\Lambda}}

\newcommand{\WW}{\mathcal{W}}
\newcommand{\JJ}{\mathcal{J}}
\newcommand{\vev}[1]{\left\langle #1 \right\rangle}
\newcommand{\vvev}[1]{\vev{\kern-0.3em\left\langle #1
    \right\rangle\kern-0.3em}}
\usepackage{amsmath}
\usepackage{amsfonts}
\usepackage{graphicx}

\begin{document}

\preprint{KOBE-TH-17-02}

\title{The generating functional of correlation functions\\ as a high
  momentum limit of a Wilson action}


\author{H.~Sonoda}
\affiliation{Physics Department, Kobe University, Kobe 657-8501, Japan}


\date{\today}

\begin{abstract}
 It is well known that a Wilson action reduces to the generating
  functional of connected correlation functions as we take the
  momentum cutoff to zero.  For a fixed point Wilson action, this
  implies that for momenta large compared with the cutoff, the action
  reduces to the generating functional.  We elaborate on this simple
  observation.
\end{abstract}


\maketitle

\section{Introduction}

A Wilson action is usually thought of as a functional integral half
done: the field with momenta below the ultraviolet (UV) cutoff still
needs to be integrated \cite{Wilson:1973jj}. To obtain full
correlation functions from a Wilson action, we have two ways.  We can
compute the correlation functions by functional integration over the
exponentiated Wilson action.  Thanks to the UV momentum cutoff
incorporated into the action, this functional integration is well
defined.  Alternatively, we can lower the momentum cutoff all the way
to zero, where the Wilson action becomes the generating functional of
the connected correlation functions.  The two ways are equivalent, and
neither is easy.

The exception is given by scale invariant theories.  In the
dimensionless convention of the renormalization group (RG), where the
momentum cutoff stays fixed, the scale invariant theories correspond
to fixed points under the RG transformation.  Given a fixed point
Wilson action (getting it is actually the hard part), we can switch to
the dimensionful convention, where the momentum cutoff $\Lambda$
decreases under the RG flow.  The Wilson action now depends on
$\Lambda$, but the dependence is given by simple scaling.  It is
trivial to take $\Lambda$ to zero, obtaining the correlation
functions.  Transcribing the vanishing cutoff limit into the
dimensionless convention, the correlation functions appear as a high
momentum limit of the Wilson action because any finite momentum in
units of the vanishing cutoff becomes large.

Considering how simple the idea is, the reader may find the paper too
long or even unnecessary.  Our excuse is that the purpose of the paper
is to provide a technically robust derivation to justify the idea.  We
use the formalism of the exact renormalization group (ERG) for generic
real scalar theories in $D$-dimensional Euclidean space (Sec.~11 of
\cite{Wilson:1973jj}).

To reach a wide range of readers including those who have not been
much exposed to the ERG formalism, we have provided plenty of
background materials.  In fact most of what is written here can be
considered a review.  To derive the main result of the paper, which is
Eq.~(\ref{mainresult}), all we have to do is to collect the right
background materials and present them in the right order.  It is
helpful if the reader is familiar with the idea of ERG through the
reading of the first third (up to Eq.~(19)) of
\cite{Polchinski:1983gv}.

We organize the paper as follows.  In Sec.~\ref{review} we review ERG
by following the perturbative treatment of \cite{Polchinski:1983gv}.
The goal of this section is to introduce the idea of a generating
functional $W_\Lambda$ with an infrared (IR) cutoff $\Lambda$, and to
show that it becomes the generating functional $\WW$ of the connected
correlation functions in the limit that $\Lambda$ goes to zero.  In
Sections \ref{generalization} and \ref{anomalous-dimension} we
generalize the ERG formalism just enough for the discussion of fixed
points in Sec.~\ref{fixed-points}, where we derive the main result
(\ref{mainresult}) that gives the connected correlation functions of a
fixed point theory as a high momentum limit of its Wilson action.
Sec.~\ref{fixed-points} is followed by two short sections: in
Sec.~\ref{section-invariance} we check the consistency of
(\ref{mainresult}) with potential conformal invariance, and in
Sec.~\ref{section-massive} we extend (\ref{mainresult}) to massive
theories.  We conclude the paper in Sec.~\ref{conclusion}.  We have
prepared three appendices.  In Appendix \ref{appendix-diffusion} we
show how to derive the diffusion equation satisfied by the generating
functional with an IR cutoff, starting from the ERG differential
equation of the corresponding Wilson action.  In Appendix
\ref{appendix-conversion} we give details of conversion between the
dimensionless and dimensionful conventions.  In Appendix
\ref{appendix-Gamma} we rewrite (\ref{mainresult}) for the effective
action.  Throughout the paper we use the shorthand notation such as
\begin{equation}
\int_p = \int \frac{d^D p}{(2 \pi)^D},\quad
\delta (p) = (2 \pi)^D \delta^{(D)} (p)\,.
\end{equation}

\section{Review\label{review}}

We review Wilson's ERG (Sec.~11 of \cite{Wilson:1973jj}) following the
perturbative treatment by J.~Polchinski \cite{Polchinski:1983gv}.  We
rely on perturbation theory for intuition, but the results we review
below should be valid beyond perturbation theory.

We consider the action
\begin{equation}
\SL [\phi] = - \frac{1}{2} \int_p \frac{p^2 + m^2}{K(p/\Lambda)} \phi
(p) \phi (-p) + \SIL [\phi]\,,
\end{equation}
where $\SIL$ consists of interaction vertices.  The free part of the
action gives the propagator
\begin{equation}
\frac{K(p/\Lambda)}{p^2 + m^2}\,,\label{K-prop}
\end{equation}
where $K(p/\Lambda)$ is a decreasing positive function of
$p^2/\Lambda^2$ such as
\begin{equation}
K (p/\Lambda) = e^{- \frac{p^2}{\Lambda^2}}\,.
\end{equation}
If $K(p/\Lambda)$ decays fast enough for large $p^2 > \Lambda^2$, and
if the interaction part is reasonable, the theory defined by
$S_\Lambda$ is free of UV divergences.  We can regard $\Lambda$ as the
UV cutoff of the theory.  Thus, we can assume that the correlation
functions given by functional integrals
\begin{equation}
\vev{\phi (p_1) \cdots \phi (p_n)}_{\SL} = \int [d\phi]\, e^{\SL [\phi]}
\phi (p_1) \cdots \phi (p_n)
\end{equation}
are well defined.  (Note that we use $e^{\SL}$ rather than $e^{-\SL}$
as the weight of integration.)

Now, how do we determine the $\Lambda$-dependence of $\SIL$?  The
short answer is that we give such $\Lambda$-dependence of $\SIL$ that
compensates the $\Lambda$-dependence of the propagator.  Let us
elaborate on this.  When we lower the cutoff infinitesimally from
$\Lambda$ to $\Lambda\, e^{-\Delta t} < \Lambda$, the propagator
changes by
\begin{equation}
- \Delta t \frac{\Lambda \frac{\partial}{\partial \Lambda}
  K(p/\Lambda)}{p^2 + m^2}\,.\label{propagator-change}
\end{equation}
The functional integrals using the same interaction part $\SIL$ change
accordingly.  If we wish to keep the same functional integrals, we
must change the interaction part to compensate the effect of
(\ref{propagator-change}).  The required compensation comes in two
types: two vertices connected by minus (\ref{propagator-change}) and
single vertices with a loop given by minus (\ref{propagator-change}).
This results in the differential equation
\begin{equation}
- \Lambda \frac{\partial}{\partial \Lambda} \SIL = \int_p
\frac{\Lambda \frac{\partial}{\partial \Lambda} K(p/\Lambda)}{p^2+m^2}
\frac{1}{2} \lb \frac{\delta \SIL}{\delta \phi (p)} \frac{\delta
  \SIL}{\delta \phi (-p)} + \frac{\delta^2 \SIL}{\delta \phi (p)
  \delta \phi (-p)}\rb\,.\label{polchinski}
\end{equation}
Exponentiating $\SIL$, we can rewrite this as
\begin{equation}
- \Lambda \frac{\partial}{\partial \Lambda} e^{\SIL} = \int_p
\frac{\Lambda \frac{\partial}{\partial \Lambda} K(p/\Lambda)}{p^2+m^2}
\frac{1}{2} \frac{\delta^2}{\delta \phi (p)
  \delta \phi (-p)} \,e^{\SIL}\,,
\end{equation}
which is a functional generalization of the diffusion equation.

Hence, as far as the internal propagators go, their cutoff dependence is
compensated by the cutoff dependence of $\SIL$.  But the external
lines still depend on $\Lambda$, and the two-point function and the
connected part of the higher-point functions acquire the following
$\Lambda$-dependence:
\begin{subequations}
\begin{eqnarray}
\vev{\phi (p) \phi (q)}_{S_\Lambda} &=& 
  \frac{K(p/\Lambda)}{p^2+m^2} \delta (p+q) +
  \frac{K(p/\Lambda)}{p^2+m^2} G_2 (p,q) 
  \frac{K(q/\Lambda)}{q^2+m^2} \,,\\
\vev{\phi (p_1) \cdots \phi (p_n)}_{S_\Lambda}^{\mathrm{connected}}
&=& \left(\prod_{i=1}^n \frac{K(p_i/\Lambda)}{p_i^2+m^2} \right) \cdot
G^{\mathrm{connected}} _n (p_1, \cdots, p_n) \,,\quad (n > 2)
\end{eqnarray}
\end{subequations}
where both $G_2$ and $G^{\mathrm{connected}}_{n > 2}$ correspond to
the sums of diagrams with the amputated external lines, and they are
independent of the cutoff $\Lambda$.  ($G_2 (p,q)$ is
proportional to $\delta (p+q)$ in the absence of symmetry breaking.)

To extract the $\Lambda$-independent correlation functions, we must
remove the cutoff functions from the external lines.  For the
connected part of the higher point functions, we can do this simply by
factoring out the cutoff functions:
\begin{eqnarray}
\vvev{\phi (p_1) \cdots \phi (p_n)}^{\mathrm{connected}}
&\equiv& \prod_{i=1}^n \frac{1}{K(p_i/\Lambda)} \cdot \vev{\phi (p_1) \cdots
  \phi (p_n)}_{S_\Lambda}^{\mathrm{connected}} \\
&=& \prod_{i=1}^n \frac{1}{p_i^2+m^2} \cdot G_n^{\mathrm{connected}}
(p_1, \cdots, p_n)\nn\,.
\end{eqnarray}
For the two-point function, we first subtract a high momentum
propagator to get
\begin{eqnarray}
&&\vev{\phi (p) \phi (q)}_{S_\Lambda} - \frac{K(p/\Lambda)\left(1 -
    K(p/\Lambda)\right)}{p^2+m^2} \delta (p+q)\nn\\
&&\quad = K(p/\Lambda)^2 \left( \frac{1}{p^2+m^2}  \delta (p+q) +
  \frac{1}{p^2+m^2} G_2 (p,q) \frac{1}{q^2+m^2}\right)\,.
\end{eqnarray}
Then, factoring out the cutoff function, we obtain a
$\Lambda$-independent two-point function:
\begin{eqnarray}
\vvev{\phi (p) \phi (q)} &\equiv& \frac{1}{K(p/\Lambda)^2}
\left( \vev{\phi (p) \phi (q)}_{S_\Lambda} - \frac{K(p/\Lambda)\left(1 -
    K(p/\Lambda)\right)}{p^2 + m^2} \delta (p+q) \right)\\
&=&  \frac{1}{p^2+m^2} \delta (p+q) + \frac{1}{p^2+m^2} G_2 (p,q)
  \frac{1}{q^2+m^2} \,.\nn
\end{eqnarray}
Incorporating the disconnected part, we can express the full
correlation functions as
\begin{eqnarray}
&&\vvev{\phi (p_1) \cdots \phi (p_n)}
\equiv \prod_{i=1}^n \frac{1}{K(p_i/\Lambda)}\nn\\
&&\quad \times 
\vev{\exp \left( - \frac{1}{2} \int_p \frac{K(p/\Lambda)\left(1 -
        K(p/\Lambda)\right)}{p^2+m^2} \frac{\delta^2}{\delta \phi (p)
      \delta \phi (-p)} \right) \phi (p_1) \cdots \phi
  (p_n)}_{S_\Lambda}\,.\label{Lambda-ind}
\end{eqnarray}
To show how the exponentiated double differentiation works, we give an
example of the four-point function:
\begin{eqnarray}
  &&\vvev{\phi (p_1) \phi (p_2) \phi (p_3) \phi (p_4)}\nn\\
  &&= \prod_{i=1}^4 \frac{1}{K(p_i/\Lambda)} \Big[
  \vev{\phi (p_1) \phi (p_2) \phi (p_3) \phi (p_4)}_{\SL}\nn\\
  &&\qquad -
  \frac{K(p_1/\Lambda)\left(1-K(p_1/\Lambda)\right)}{p_1^2+m^2} \delta
  (p_1+p_2) \vev{\phi (p_3) \phi (p_4)}_{\SL}\nn\\
  &&\qquad - \frac{K(p_3/\Lambda)\left(1-K(p_3/\Lambda)\right)}{p_3^2+m^2} \delta
  (p_3+p_4) \vev{\phi (p_1) \phi (p_2)}_{\SL}\nn\\
  &&\qquad +
  \frac{K(p_1/\Lambda)\left(1-K(p_1/\Lambda)\right)}{p_1^2+m^2}
  \delta (p_1+p_2)
  \frac{K(p_3/\Lambda)\left(1-K(p_3/\Lambda)\right)}{p_3^2+m^2} \delta 
  (p_3+p_4)\nn\\
  &&\quad + (\textrm{t, u-channels}) \, \Big]\,.
\end{eqnarray}
It is commonly taken for granted that only the low momentum
correlation functions are kept invariant under the exact
renormalization group transformations, but we have shown more than
that: via (\ref{Lambda-ind}) we can recover the entire cutoff
independent correlation functions. (This was first pointed out in
\cite{Sonoda:2007dj}, and has been used extensively for the
realization of symmetry in the ERG formalism \cite{Igarashi:2009tj}.)
Now, introducing a source $\JJ (p)$, and summing (\ref{Lambda-ind})
over all $n$, we can express the generating functional $\WW [\JJ]$ of
the connected correlation functions as
\begin{eqnarray}
e^{\WW [\JJ]} &\equiv& \sum_{n=0}^\infty \frac{1}{n!} \int_{p_1,
                       \cdots, p_n} \JJ (-p_1) \cdots \JJ (-p_n)
                       \vvev{\phi (p_1) \cdots \phi (p_n)}\nn\\
&=& \vvev{\exp \left( \int_p \JJ (-p) \phi (p) \right)} \nn\\
&=& \vev{\exp \left( - \frac{1}{2} \int_p
    \frac{K(p/\Lambda)\left(1-K(p/\Lambda)\right)}{p^2+m^2}
\frac{\delta^2}{\delta \phi (p) \delta \phi (-p)}\right)
    \exp \left( \int_p \frac{\JJ (-p) \phi
    (p)}{K(p/\Lambda)}\right)}_{\SL}\nn\\
&=& \int [d\phi] \exp \Big[ \SL [\phi] \nn\\
&&\quad + \int_p \left( - \frac{1}{2} 
 \frac{K(p/\Lambda)\left(1-K(p/\Lambda)\right)}{p^2+m^2} \frac{\JJ
    (-p) \JJ (p)}{K(p/\Lambda)^2} + \frac{\JJ (-p) \phi
    (p)}{K(p/\Lambda)} \right) \Big]\,. \label{WW-def}
\end{eqnarray}

So far we have regarded $\Lambda$ as a UV cutoff because the momentum
modes with $p > \Lambda$ are suppressed in the functional integration
over $e^{\SL}$.  Let us note, however, that the interaction part
$\SIL$ results from integrating over the modes with momentum higher
than $\Lambda$.  So, if we regard $\SIL$ as a consequence of
functional integration, we may call $\Lambda$ an IR cutoff.  Since the
propagator of the high momentum modes is
\begin{equation}
\frac{1 - K(p/\Lambda)}{p^2 + m^2}\,,
\end{equation}
the generating functional of the connected correlation functions with
an IR cutoff $\Lambda$ is defined by
\begin{equation}
W_\Lambda [J] \equiv \frac{1}{2} \int_p \frac{1-K(p/\Lambda)}{p^2+m^2}
J(p) J(-p) + \SIL \left[ \frac{1-K(p/\Lambda)}{p^2+m^2}
  J(p)\right]\,, \label{WL-def}
\end{equation}
where we have added the free part, and substituted
\begin{equation}
\phi (p) = \frac{1-K(p/\Lambda)}{p^2+m^2} J(p)
\end{equation}
into the interaction part.  ($W_\Lambda$ was first introduced in
\cite{Morris:1994ie}.)  Using the full action, we can rewrite
$W_\Lambda$ as
\begin{equation}
W_\Lambda [J] = \frac{1}{2} \int_p \frac{J (p) J(-p)}{R_\Lambda (p)}
 + \SL \left[ \frac{1-K(p/\Lambda)}{p^2+m^2}
  J(p)\right] \,, \label{WL-SL}
\end{equation}
where
\begin{equation}
R_\Lambda (p) \equiv \frac{K(p/\Lambda)}{1-K(p/\Lambda)} (p^2+m^2)
\label{RL-K}
\end{equation}
is a positive cutoff function that decays rapidly for $p^2 >
\Lambda^2$.  We note
\begin{equation}
\lim_{\Lambda \to 0+} R_\Lambda (p) = 0\,.\label{vanishing-RL}
\end{equation}
Now, using $W_\Lambda [J]$ instead of $\SL [\phi]$ in (\ref{WW-def}),
and using $J$ instead of $\phi$ as integration variables, we obtain a
simpler expression for $\WW$:
\begin{eqnarray}
e^{\WW [\JJ]} &=& \int [d\phi] \exp \left[ \SL [\phi] + \int_p \left(
    - \frac{1}{2} \frac{\JJ (p) \JJ (-p)}{R_\Lambda (p)} + \frac{\JJ
      (-p) \phi (p)}{K(p/\Lambda)} \right)\right]\nn\\
&=& \int [dJ] \exp \left[ W_\Lambda [J] - \frac{1}{2} \int_p
  \frac{1}{R_\Lambda (p)} \left( J(p) - \JJ (p) \right)\left( J (-p) -
    \JJ (-p) \right)\right]\,.\label{WW-W}
\end{eqnarray}

It is straightforward to obtain the cutoff dependence of $W_\Lambda
[J]$.  Since $W_\Lambda [J]$ is defined by (\ref{WL-def}), and the
$\Lambda$-dependence of $\SIL$ is given by (\ref{polchinski}), we
obtain
\begin{subequations}
\label{polchinski-diffusion}
\begin{equation}
- \Lambda \frac{\partial}{\partial \Lambda} W_\Lambda [J] =
\frac{1}{2} \int_p \Lambda \frac{\partial R_\Lambda (p)}{\partial
  \Lambda} \lb \frac{\delta W_\Lambda [J]}{\delta J(-p)} \frac{\delta
  W_\Lambda [J]}{\delta J(p)} +  \frac{\delta^2 W_\Lambda [J]}{\delta
  J(-p) \delta J(p)} \rb\,,
\end{equation}
or equivalently
\begin{equation}
- \Lambda \frac{\partial}{\partial \Lambda} e^{W_\Lambda [J]}
= \int_p \Lambda \frac{\partial R_\Lambda (p)}{\partial
  \Lambda} \frac{1}{2} \frac{\delta^2}{\delta J (p) \delta J(-p)}
e^{W_\Lambda [J]}\,.
\end{equation}
\end{subequations}
(See Appendix \ref{appendix-diffusion} for derivation.)  This
functional diffusion equation can be solved formally: for $\Lambda' <
\Lambda$, we obtain
\begin{eqnarray}
e^{W_{\Lambda'} [J]} &=& \exp \left( \int_p \left(R_\Lambda (p) -
    R_{\Lambda'} (p)\right) \frac{1}{2} \frac{\delta^2}{\delta J(p)
    \delta J(-p)} \right) e^{W_\Lambda [J]}\nn\\
&=& \int [dJ'] \exp \Big[ W_\Lambda [J']\nn\\
&&\quad- \frac{1}{2} \int_p \frac{1}{R_\Lambda (p) - R_{\Lambda'} (p)} \left(
  J' (p) - J(p)\right) \left( J' (-p) - J(-p)\right)\Big]\,.
\label{WL-solution}
\end{eqnarray}
Comparing this with (\ref{WW-W}) and using (\ref{vanishing-RL}),
we obtain
\begin{equation}
\WW [\JJ] = \lim_{\Lambda \to 0+} W_\Lambda [\JJ]\,.\label{polchinski-WW-WL}
\end{equation}
This is the well known equality referred to at the beginning of the
abstract of the paper.  Since $W_\Lambda$ is directly related to $\SL$
by (\ref{WL-SL}), we can say that Eq.~(\ref{polchinski-WW-WL}) gives
the generating functional of the connected correlation functions as
the zero cutoff limit of the Wilson action.

\section{Generalization\label{generalization}}

In the previous section we have summarized Wilson's ERG following
\cite{Polchinski:1983gv}.  We have introduced two types of generating
functionals: $W_\Lambda$ with an IR cutoff and $\WW$ without.  In this
section we would like to generalize the formalism in two ways.  So
far, we have introduced only one cutoff function $K(p/\Lambda)$.
Another cutoff function $R_\Lambda (p)$ is given in terms of
$K(p/\Lambda)$ by (\ref{RL-K}).  Our first generalization follows
\cite{Sonoda:2015bla}, and we introduce $K_\Lambda (p)$ and $R_\Lambda
(p)$ as two independent positive cutoff functions. (This is necessary
not only for the second generalization, but also if we wish to include
the original formulation of \cite{Wilson:1973jj} under the same
footing.)  The second generalization, to be introduced in the next
section, follows \cite{Igarashi:2016qdr}, and we introduce an
anomalous dimension to the scalar field.

Using two independent cutoff functions, we define the correlation
functions by
\begin{eqnarray}
&&\vvev{\phi (p_1) \cdots \phi (p_n)}_{S_\Lambda}^{K_\Lambda, R_\Lambda}
\equiv \prod_{i=1}^n
\frac{1}{K_\Lambda (p_i)}\nn\\
&&\quad \times 
\vev{\exp \left( - \frac{1}{2} \int_p \frac{K_\Lambda (p)^2}{R_\Lambda (p)}
        \frac{\delta^2}{\delta \phi (p)
      \delta \phi (-p)} \right) \phi (p_1) \cdots \phi
  (p_n)}_{S_\Lambda}\,.\label{mod-corr}
\end{eqnarray}
In the previous section we have chosen
\begin{equation}
K_\Lambda (p) = K(p/\Lambda),\quad
R_\Lambda (p) = \frac{K(p/\Lambda)}{1-K(p/\Lambda)} (p^2+m^2)\,,
\label{polchinski-KR}
\end{equation}
for which (\ref{mod-corr}) reduces to (\ref{Lambda-ind}).  We assume
in general that both $K_\Lambda$ and $R_\Lambda$ decay rapidly for
large momenta $p^2 > \Lambda^2$.  This implies
\begin{equation}
\lim_{\Lambda \to 0+} R_\Lambda (p) = 0\,.\label{vanishing-RL-general}
\end{equation}

For (\ref{mod-corr}) to be independent of $\Lambda$, the Wilson action
must satisfy
\begin{eqnarray}
- \Lambda \frac{\partial}{\partial \Lambda} \SL &=& \int_p \Lambda
\frac{\partial}{\partial \Lambda} \ln K_\Lambda (p) \cdot \phi (p)
\frac{\delta \SL}{\delta \phi (p)}\nn\\
&& + \int_p \Lambda \frac{\partial}{\partial \Lambda} R_\Lambda (p)
\cdot \frac{K_\Lambda (p)^2}{R_\Lambda (p)^2} \frac{1}{2} \lb
\frac{\delta \SL}{\delta \phi (p)} \frac{\delta \SL}{\delta \phi (-p)}
+ \frac{\delta^2 \SL}{\delta \phi (p) \delta \phi (-p)} \rb\,.
\label{ERGdiffeq}
\end{eqnarray}
For the choice (\ref{polchinski-KR}), this reduces to
\begin{eqnarray}
- \Lambda \frac{\partial}{\partial \Lambda} \SL &=& \int_p \Lambda
\frac{\partial}{\partial \Lambda} \ln K(p/\Lambda) \cdot \phi (p)
\frac{\delta \SL}{\delta \phi (p)}\nn\\
&& + \int_p \frac{\Lambda \frac{\partial}{\partial \Lambda}
  K(p/\Lambda)}{p^2+m^2} \frac{1}{2} \lb 
\frac{\delta \SL}{\delta \phi (p)} \frac{\delta \SL}{\delta \phi (-p)}
+ \frac{\delta^2 \SL}{\delta \phi (p) \delta \phi (-p)} \rb\,,
\end{eqnarray}
which is (\ref{polchinski}) rewritten for the total action.  We do not
derive (\ref{ERGdiffeq}) here; we refer the interested reader to
\cite{Sonoda:2015bla} for derivation.

Now, we define the generating functional with an IR cutoff in the same
way as before by
\begin{subequations}
\label{WL-J-def}
\begin{equation}
W_\Lambda [J] \equiv \frac{1}{2} \int_p \frac{J (p) J(-p)}{R_\Lambda (p)} +
\SL [\phi]\,,
\end{equation}
where
\begin{equation}
J(p) \equiv \frac{R_\Lambda (p)}{K_\Lambda (p)} \phi (p)\,.
\end{equation}
\end{subequations}
For the choice (\ref{polchinski-KR}), the above reduces to
(\ref{WL-SL}) and (\ref{RL-K}). Using (\ref{ERGdiffeq}), it is
straightforward to show that $W_\Lambda [J]$ satisfies the same
equation as (\ref{polchinski-diffusion}):
\begin{equation}
- \Lambda \frac{\partial}{\partial \Lambda} e^{W_\Lambda [J]} = \int_p
\Lambda \frac{\partial R_\Lambda (p)}{\partial \Lambda} \frac{1}{2}
\frac{\delta^2}{\delta J(p) \delta J(-p)} e^{W_\Lambda [J]}\,.
\label{diffusion}
\end{equation}
(See Appendix \ref{appendix-diffusion} for derivation.)  The rest
proceeds the same way as in the previous section.  The generating
functional of the connected correlation functions, defined by
\begin{equation}
e^{\WW [\JJ]} \equiv \vvev{\exp \left( \int_p \JJ (-p) \phi (p)
  \right)}_{\SL}^{K_\Lambda, R_\Lambda}\,,
\end{equation}
is given by
\begin{equation}
e^{\WW [\JJ]} = \int [dJ] \exp \left( W_\Lambda [J] - \frac{1}{2}
  \int_p \frac{1}{R_\Lambda (p)} \left( J(p) - \mathcal{J} (p) \right)
  \left( J(-p) - \mathcal{J} (-p) \right)\right)\,.
\end{equation}
Hence,  we obtain the same result as
(\ref{polchinski-WW-WL}):
\begin{equation}
\WW [\JJ] = \lim_{\Lambda \to 0+} W_\Lambda [\JJ]\,,\label{WW-WL}
\end{equation}
where we have used (\ref{vanishing-RL-general}).

\section{Anomalous dimension\label{anomalous-dimension}}

In this section, we introduce an anomalous dimension of the scalar
field.  A nonvanishing anomalous dimension is required by the
nontrivial fixed point to be discussed in the next section.  Let $\SL$
be the Wilson action for which
\begin{equation}
\vvev{\phi (p_1) \cdots \phi (p_n)}_{\SL}^{K_\Lambda, R_\Lambda}
\end{equation}
are independent of $\Lambda$.  We wish to construct
$\Lambda$-dependent Wilson actions $\tilde{S}_\Lambda$ so that
\begin{equation}
\vvev{\phi (p_1) \cdots \phi (p_n)}_{\tilde{S}_\Lambda}^{K_\Lambda,
  R_\Lambda}
= \left(\frac{\mu}{\Lambda}\right)^{n \gamma} \vvev{\phi (p_1) \cdots
  \phi (p_n)}_{S_\mu}^{K_\mu,  R_\mu}\,.
\label{StildeL-SL}
\end{equation}
Here, $\mu$ is a fixed reference scale chosen arbitrarily.  For
simplicity, we have chosen the anomalous dimension $\gamma$ as a
constant independent of $\Lambda$.  At $\Lambda = \mu$, 
the two actions agree:
\begin{equation}
\tilde{S}_\mu = S_\mu\,.
\end{equation}
Unlike $\SL$, the correlation functions of $\tilde{S}_\Lambda$ are
$\Lambda$-dependent, but the $\Lambda$-dependence is merely a change
of normalization of the field.  We wish to relate $\tilde{S}_\Lambda$
to $\SL$ in the following.

We rewrite (\ref{StildeL-SL}) as
\begin{eqnarray}
\vvev{\phi (p_1) \cdots \phi
  (p_n)}_{\tilde{S}_\Lambda}^{\left(\frac{\mu}{\Lambda}\right)^\gamma
  K_\Lambda, \left(\frac{\mu}{\Lambda}\right)^{2\gamma} R_\Lambda}
&=& \left(\frac{\Lambda}{\mu}\right)^{n\gamma} \vvev{\phi (p_1) \cdots
  \phi (p_n)}_{\tilde{S}_\Lambda}^{K_\Lambda, R_\Lambda}\nn\\
&=& \vvev{\phi (p_1) \cdots \phi (p_n)}_{S_\mu}^{K_\mu, R_\mu}\,.
\end{eqnarray}
Since this is independent of $\Lambda$, $\tilde{S}_\Lambda$ must
satisfy
\begin{eqnarray}
&&- \Lambda \frac{\partial}{\partial \Lambda} \tilde{S}_\Lambda [\phi]
= \int_p \left( \Lambda \frac{\partial}{\partial \Lambda} \ln
  K_\Lambda (p) - \gamma \right) \phi (p) \frac{\delta
  \tilde{S}_\Lambda}{\delta \phi (p)}\nn\\
&&\quad + \int_p \left( \Lambda \frac{\partial}{\partial \Lambda} R_\Lambda
  (p) - 2 \gamma R_\Lambda (p)\right) \frac{K_\Lambda (p)^2}{R_\Lambda
  (p)^2} \frac{1}{2} \lb \frac{\delta \tilde{S}_\Lambda}{\delta \phi
  (-p)} \frac{\delta \tilde{S}_\Lambda}{\delta \phi (p)}
+ \frac{\delta^2 \tilde{S}_\Lambda}{\delta \phi (p)\delta \phi (-p)}
\rb\,.\label{ERGdiffeq-gamma}
\end{eqnarray}
(We obtain this from (\ref{ERGdiffeq}) by substituting
$\left(\frac{\mu}{\Lambda}\right)^\gamma K_\Lambda$ and
$\left(\frac{\mu}{\Lambda}\right)^{2\gamma} R_\Lambda$ into
$K_\Lambda$ and $R_\Lambda$, respectively.)  We define the generating
functional $\tilde{W}_\Lambda$ with an IR cutoff for
$\tilde{S}_\Lambda$, using the same cutoff functions as for $\SL$:
\begin{equation}
\tilde{W}_\Lambda [J] \equiv \frac{1}{2} \int_p \frac{J(p)
  J(-p)}{R_\Lambda (p)} + \tilde{S}_\Lambda [\phi]\,,
\end{equation}
where
\begin{equation}
J(p) \equiv \frac{R_\Lambda (p)}{K_\Lambda (p)} \phi (p)\,.
\end{equation}
Using (\ref{ERGdiffeq-gamma}), we can derive the cutoff dependence of
$\tilde{W}_\Lambda$ as
\begin{equation}
- \Lambda \frac{\partial}{\partial \Lambda} e^{\tilde{W}_\Lambda [J]}
= \int_p \left[ \gamma J(p) \frac{\delta}{\delta J(p)} + \left(
    \Lambda \frac{\partial R_\Lambda (p)}{\partial \Lambda} - 2 \gamma
    R_\Lambda (p) \right) \frac{1}{2} \frac{\delta^2}{\delta J(p)
    \delta J(-p)} \right] e^{\tilde{W}_\Lambda [J]}\,.\label{diffusion-gamma} 
\end{equation}
(See Appendix \ref{appendix-diffusion} for derivation.)  To solve this
under the initial condition
\begin{equation}
\tilde{W}_\mu [J] = W_\mu [J]\,,
\end{equation}
we first rewrite the equation as
\begin{eqnarray}
&&- \Lambda \frac{\partial}{\partial \Lambda} 
\exp \left( \tilde{W}_\Lambda \left[
    \left(\frac{\Lambda}{\mu}\right)^\gamma J \right] \right)\nn\\
&=& \int_p \Lambda \frac{\partial}{\partial \Lambda} \left(
\left(\frac{\Lambda}{\mu}\right)^{-2\gamma} R_\Lambda (p)\right) 
\frac{1}{2} \frac{\delta^2}{\delta J(p) \delta J(-p)} \exp \left(
  \tilde{W}_\Lambda \left[ 
    \left(\frac{\Lambda}{\mu}\right)^\gamma J \right] \right)\,.
\end{eqnarray}
This is solved by
\begin{eqnarray}
&&\exp \left( \tilde{W}_\Lambda \left[
    \left(\frac{\Lambda}{\mu}\right)^\gamma J \right] \right)\nn\\
&=& \exp \left[ \int_p \left( R_\mu (p) -
    \left(\frac{\Lambda}{\mu}\right)^{-2\gamma} R_\Lambda (p) \right)
  \frac{1}{2} \frac{\delta^2}{\delta J(p) \delta J(-p)} \right]
e^{W_\mu [J]}\,.
\end{eqnarray}

To relate $W_\Lambda$ to $\tilde{W}_\Lambda$, we compare the above
solution with
\begin{equation}
e^{W_\Lambda [J]} = \exp \left[ \int_p \left(R_\mu (p) - R_\Lambda
    (p)\right) \frac{1}{2} \frac{\delta^2}{\delta J(p) \delta J(-p)}
\right] e^{W_\mu [J]}\,,
\end{equation}
which is obtained from the first line of (\ref{WL-solution}).  We
easily obtain
\begin{equation}
e^{W_\Lambda [J]} =
\exp \left[ \left(\left(\frac{\Lambda}{\mu}\right)^{-2\gamma} - 1
  \right) \int_p R_\Lambda (p) \frac{1}{2} \frac{\delta^2}{\delta J(p)
    \delta J(-p)} \right] \exp \left( \tilde{W}_\Lambda \left[
    \left(\frac{\Lambda}{\mu}\right)^\gamma J \right] \right)\,.
\label{WL-WLtilde}
\end{equation}
We could rewrite this as a relation between $\SL$ and
$\tilde{S}_\Lambda$, but we do not need it.

We end this section by giving $\WW [\JJ]$ as a limit of
$\tilde{W}_\Lambda$.  We assume
\begin{equation}
\gamma > 0
\end{equation}
so that $(\Lambda/\mu)^{-2\gamma}$ dominates over $1$ as $\Lambda \to 0+$.
If we assume further
\begin{equation}
\lim_{\Lambda \to 0+} \Lambda^{-2\gamma} R_\Lambda (p) =
0\,,\label{vanishing-RL-gamma} 
\end{equation}
which is a little stronger than (\ref{vanishing-RL-general}), we
obtain from (\ref{WW-WL}) and (\ref{WL-WLtilde})
\begin{equation}
\WW [\JJ] = \lim_{\Lambda \to 0+} W_\Lambda [\JJ] = \lim_{\Lambda \to 0+}
\tilde{W}_\Lambda \left[ 
    \left(\frac{\Lambda}{\mu}\right)^\gamma \JJ \right]\,.
\label{WW-Wtilde}
\end{equation}

\section{Fixed points \label{fixed-points}}

The differential equation (\ref{ERGdiffeq-gamma}) or equivalently
(\ref{diffusion-gamma}) does not have a fixed point solution for an
obvious reason: the cutoff $\Lambda$ keeps changing.  We need to adopt
the dimensionless convention in which we measure all the physical
quantities in units of appropriate powers of the cutoff $\Lambda$.  We
give a table of conversion with the dimensionful quantities on the
left, and the corresponding dimensionless quantities (with bars except
for $K, R$) on the right:
\begin{subequations}
\label{table-conversion}
\begin{eqnarray}
\Lambda &=& \mu e^{-t}\,,\\
\phi (p) &=& \Lambda^{-\frac{D+2}{2}} \bar{\phi} (p/\Lambda)\,,\\
J(p) &=& \Lambda^{-\frac{D-2}{2}} \bar{J} (p/\Lambda)\,,\label{JJbar}\\
K_\Lambda (p) &=& K(p/\Lambda)\,,\\
R_\Lambda (p) &=& \Lambda^2 R (p/\Lambda)\,,\\
\tilde{S}_\Lambda [\phi] &=& \bar{S}_t [\bar{\phi}]\,,\\
\tilde{W}_\Lambda [J] &=& \bar{W}_t [\bar{J}]\,.
\end{eqnarray}
\end{subequations}
We assume that the dimensionless cutoff functions satisfy
\begin{equation}
\lim_{\Lambda \to 0+} K(p/\Lambda) = \lim_{\Lambda \to 0+} R
(p/\Lambda) = 0\,.
\end{equation}
Hence, if the anomalous dimension satisfies
\begin{equation}
0 \le \gamma \le 1\,,
\end{equation}
we obtain (\ref{vanishing-RL-gamma}).

The correlation functions in the dimensionless convention are related
to those in the dimensionful convention by
\begin{subequations}
\begin{equation}
\vvev{\bar{\phi} (p_1) \cdots \bar{\phi} (p_n)}_{\bar{S}_t}^{K, R}
= \Lambda^{n \frac{D+2}{2}} \vvev{\phi (p_1 \Lambda) \cdots \phi
  (p_n \Lambda)}_{\tilde{S}_\Lambda}^{K_\Lambda, R_\Lambda}\,.
\end{equation}
Using (\ref{StildeL-SL}), we can rewrite the right-hand side using
$\Lambda$-independent correlation functions as
\begin{equation}
\vvev{\bar{\phi} (p_1) \cdots \bar{\phi} (p_n)}_{\bar{S}_t}^{K, R}
= \Lambda^{n \frac{D+2}{2}} \left(\frac{\mu}{\Lambda}\right)^{n
  \gamma} \vvev{\phi (p_1 \Lambda) \cdots \phi (p_n
  \Lambda)}_{\SL}^{K_\Lambda, R_\Lambda}\,.
\end{equation}
\end{subequations}
Hence, in the dimensionless convention the correlation functions
satisfy the following scaling relation
\begin{equation}
\vvev{\bar{\phi} (p_1 e^{\Delta t}) \cdots \bar{\phi} (p_n e^{\Delta
    t})}_{\bar{S}_{t+\Delta t}}^{K, R} 
= \exp \left( n \left(- \frac{D+2}{2} + \gamma \right) \Delta t\right) 
\vvev{\bar{\phi} (p_1) \cdots \bar{\phi} (p_n)}_{\bar{S}_t}^{K,R}\,.
\label{scaling}
\end{equation}
Note that we are comparing the correlation functions for different
sets of momenta at two different Wilson actions which are related by
ERG.

It is straightforward to obtain the ERG differential equations for
$\bar{S}_t$ and $\bar{W}_t$ by rewriting the equations for
$\tilde{S}_\Lambda$ and $\tilde{W}_\Lambda$.  For the rewriting we use
\begin{eqnarray}
\frac{\delta}{\delta \phi (p)} 
&=& \int_q \frac{\delta \bar{\phi} (q)}{\delta \phi (p)}
    \frac{\delta}{\delta \bar{\phi} (q)}
= \int_q \frac{\delta \left(\Lambda^{\frac{D+2}{2}} \phi (q
    \Lambda)\right)}{\delta \phi (p)}  \frac{\delta}{\delta \bar{\phi}
    (q)}\nn\\
&=& \int_q \Lambda^{\frac{D+2}{2}} \delta (q \Lambda - p)
\frac{\delta}{\delta \bar{\phi} (q)}
= \Lambda^{- \frac{D-2}{2}} \frac{\delta}{\delta \bar{\phi}
  (p/\Lambda)}
\label{ddphi-ddphibar}
\end{eqnarray}
and the analogous
\begin{equation}
\frac{\delta}{\delta J (p)} = \Lambda^{-\frac{D+2}{2}}
\frac{\delta}{\delta \bar{J} (p/\Lambda)}\,.
\label{ddJ-ddJbar}
\end{equation}
We need only the equation for $\bar{W}_t$ here
(see Appendix \ref{appendix-conversion} for derivation):
\begin{eqnarray}
\partial_t e^{\bar{W}_t [\bar{J}]} &=& \int_p \left[
\left( p \cdot \partial_p + \frac{D-2}{2} + \gamma \right) \bar{J} (p)
\cdot \frac{\delta}{\delta \bar{J} (p)}\right.\nn\\
&&\quad\left. + \left( - p \cdot \partial_p + 2 - 2 \gamma \right)
  R(p) \cdot \frac{1}{2} \frac{\delta^2}{\delta \bar{J} (p) \delta
    \bar{J} (-p)} \right] e^{\bar{W}_t [\bar{J}]}\,.\label{diffusion-dimless}
\end{eqnarray}
For this to have a fixed point solution, we must choose $\gamma$
appropriately. With $\gamma = 0$, we only get the Gaussian fixed point:
\begin{equation}
\bar{W}_G [\bar{J}] = \frac{1}{2} \int_p \frac{\bar{J} (p) \bar{J}
  (-p)}{p^2 + R (p)}\,.
\end{equation}
By choosing $0 < \gamma < 1$ appropriately, we can obtain a nontrivial
fixed point $\bar{W}^* [\bar{J}]$ that satisfies
\begin{eqnarray}
0 &=& \int_p \left[
\left( p \cdot \partial_p + \frac{D-2}{2} + \gamma \right) \bar{J} (p)
\cdot \frac{\delta}{\delta \bar{J} (p)}\right.\nn\\
&&\quad\left. + \left( - p \cdot \partial_p + 2 - 2 \gamma \right)
  R(p) \cdot \frac{1}{2} \frac{\delta^2}{\delta \bar{J} (p) \delta
    \bar{J} (-p)} \right] e^{\bar{W}^* [\bar{J}]}\,.
\label{Wbarstar}
\end{eqnarray}
For the fixed point, the scaling relation (\ref{scaling}) relates the
correlation functions for the same fixed point Wilson action
$\bar{S}^*$:
\begin{equation}
\vvev{\bar{\phi} (p_1 e^{\Delta t}) \cdots \bar{\phi} (p_n e^{\Delta
    t})}_{\bar{S}^*}^{K,R} = \exp \left( n \left(-\frac{D+2}{2} + \gamma
  \right) \Delta t \right) \vvev{\bar{\phi} (p_1) \cdots \bar{\phi}
  (p_n)}_{\bar{S}^*}^{K,R}\,.
\end{equation}

Now we are ready to derive the main result of this paper.  For a
general theory, we get
\begin{equation}
\tilde{W}_\Lambda \left[ \left(\frac{\Lambda}{\mu}\right)^\gamma \JJ
\right] =  \bar{W}_{t = - \ln \Lambda/\mu} \left[ \bar{J} (p) =
  \left(\frac{\Lambda}{\mu}\right)^\gamma \Lambda^{\frac{D-2}{2}} \JJ
  (p \Lambda) \right]\,.
\end{equation}
Unless we know $\bar{W}_t$ for very large $t$, we cannot use
(\ref{WW-Wtilde}) to obtain $\WW [\JJ]$.  At a fixed point, however,
$\bar{W}_t$ does not depend on $t$, and the $\Lambda$-dependence of
$\tilde{W}_\Lambda [J]$ solely comes from the scaling of variables:
\begin{equation}
\tilde{W}_\Lambda [J] = \bar{W}^* \left[\bar{J} (p) =
  \Lambda^{\frac{D-2}{2}} J (p \Lambda)\right]\,,
\label{Wtilde-Wbar}
\end{equation}
where $\bar{W}^*$ is a fixed point functional satisfying
(\ref{Wbarstar}).  Substituting (\ref{Wtilde-Wbar}) into
(\ref{WW-Wtilde}), we obtain the main result of this paper
\begin{equation}
\WW [\JJ] = \lim_{\Lambda \to 0+} \bar{W}^* \left[ \bar{J} (p) = 
  \left(\frac{\Lambda}{\mu}\right)^\gamma \Lambda^{\frac{D-2}{2}} \JJ
  (p\Lambda) \right]\,, \label{mainresult}
\end{equation}
which gives the connected correlation functions as a high momentum
limit of $\bar{W}^*$.

In Sec.~1 we have briefly explained why we call (\ref{mainresult}) a
high momentum limit: any momentum in units of $\Lambda$ gets large as
we take $\Lambda \to 0+$.  To make this explanation more concrete,
expand $\bar{W}^*$ in powers of $\bar{J}$:
\begin{equation}
\bar{W}^*  [\bar{J}] = \sum_{n=1}^\infty \frac{1}{n!} \int_{p_1, \cdots,
  p_{n}} \bar{J}(-p_1) \cdots \bar{J}(-p_{n})\,\delta (p_1 + \cdots + p_{n})\,
w^*_{n} (p_1, \cdots, p_{n})\,.
\end{equation}
We then obtain
\begin{eqnarray}
&&  \bar{W}^* \left[ \left(\frac{\Lambda}{\mu}\right)^\gamma
  \Lambda^{\frac{D-2}{2}} \JJ (p\Lambda) \right] \nn\\
&=& \sum_{n=1}^\infty \frac{1}{n!} \int_{p_1, \cdots,
    p_{n}} \left(\frac{\Lambda}{\mu}\right)^{n \gamma}
\Lambda^{n \frac{D-2}{2}} \, \JJ (-p_1 \Lambda) \cdots \JJ (-p_n \Lambda)
\,\delta (p_1 + \cdots + p_{n})\nn\\
  &&\qquad \times w_{n} (p_1, \cdots, p_{n})\nn\\
  &=&\sum_{n=1}^\infty \frac{1}{n!} \int_{p_1, \cdots, p_{n}}
  \JJ (-p_1) \cdots \JJ (-p_{n}) \,\delta (p_1 + \cdots + p_{n})\nn\\
  &&\qquad \times  \left(\frac{\Lambda}{\mu}\right)^{n \gamma}
\Lambda^{-n \frac{D+2}{2}+D} \, w_{n} (p_1/\Lambda, \cdots, p_{n}/\Lambda)\,.
\end{eqnarray}
This implies
\begin{eqnarray}
&&\vvev{\phi (p_1) \cdots \phi (p_{n})}^{\mathrm{connected}}\nn\\
&&= \lim_{\Lambda \to 0+}
 \left(\frac{\Lambda}{\mu}\right)^{n \gamma}
\Lambda^{-n \frac{D+2}{2}+D} \cdot w_{n} (p_1/\Lambda, \cdots,
   p_{n}/\Lambda)
\, \delta (p_1 + \cdots + p_{n})\,.
\end{eqnarray}
Thus, the connected correlation functions are obtained as the high
momentum limit of $w_{n} (p_1, \cdots, p_{n})$.
Especially for $n=2$, we obtain
\begin{equation}
\vvev{\phi (p) \phi (q)} = \lim_{\Lambda \to 0+}
\left(\frac{\Lambda}{\mu}\right)^{2 \gamma} \frac{1}{\Lambda^2}\, w_2
(p/\Lambda, -p/\Lambda) \,\delta (p+q)\,.
\end{equation}
This implies
\begin{equation}
w_2 (p , - p ) \stackrel{p \to \infty}{\longrightarrow}
\textrm{const} \, \frac{1}{(p^2)^{1 - \gamma}}\,. 
\end{equation}

\section{Conformal invariance\label{section-invariance}}

We would like to discuss the invariance properties of $\bar{W}^*$ and $\WW$.
The fixed point theory has scale invariance, and we expect $\WW [\JJ]$
to have naive scale invariance
\begin{subequations}
\label{WW-scale}
\begin{equation}
\int_p \JJ (-p) D^S (p) \frac{\delta \WW [\JJ ]}{\delta \JJ (-p)}
= 0\,,
\end{equation}
where
\begin{equation}
D^S (p) \equiv - p \cdot \partial_p - \frac{D+2}{2} + \gamma
\end{equation}
\end{subequations}
is the generator of scale transformation.  This is a direct
consequence of (\ref{mainresult}); the very existence of the limit
implies (\ref{WW-scale}).

If the fixed point theory has also conformal invariance, we
expect
\begin{subequations}
\label{WW-conformal-DK}
\begin{equation}
\int_p \JJ (-p) D^K_\mu (p) \frac{\delta \WW [\JJ ]}{\delta \JJ (-p)}
= 0\,,\label{WW-conformal}
\end{equation}
where
\begin{equation}
D^K_\mu (p) \equiv - p_\nu \frac{\partial^2}{\partial p_\mu \partial
  p_\nu} + \frac{1}{2} p_\mu \frac{\partial^2}{\partial p_\nu \partial
  p_\nu} + \left(- \frac{D+2}{2} + \gamma \right)
\frac{\partial}{\partial p_\mu}\label{WW-DK}
\end{equation}
\end{subequations}
is the generator of special conformal transformation.  On the other
hand, it is known \cite{Rosten:2014oja, Delamotte:2015aaa,
  Rosten:2016nmc, Rosten:2016zap, Sonoda:2017zgl, Rosten:2017urs} that
the conformal invariance of the fixed point theory is realized as
\begin{eqnarray}
  &&\int_p \bar{J} (-p) D_\mu^K (p) \frac{\delta \bar{W}^* [\bar{J}]}{\delta
    \bar{J} (-p)}
  + \int_p \left( - p \cdot \partial_p + 2 - 2 \gamma \right) R
  (p) \nn\\
  &&\qquad\qquad \times \frac{1}{2} \int_q \delta (q-p) \frac{\partial}{\partial
    p_\mu} \lb \frac{\delta^2 \bar{W}^* [\bar{J}]}{\delta \bar{J}(p) \delta
    \bar{J} (-q)} + \frac{\delta \bar{W}^* [\bar{J}]}{\delta \bar{J}(p)}
  \frac{\delta \bar{W}^* [\bar{J}]}{\delta \bar{J}(-q)} \rb =
  0\,\label{W-conformal} 
\end{eqnarray}
in terms of the fixed point functional.  As a consistency check of
(\ref{mainresult}), we wish to use (\ref{mainresult}) to derive
(\ref{WW-conformal}) from (\ref{W-conformal}).

Substituting
\begin{equation}
\bar{J} (p) = \left(\frac{\Lambda}{\mu}\right)^\gamma
\Lambda^{\frac{D-2}{2}} \JJ (p \Lambda)
\end{equation}
into (\ref{W-conformal}), and using
\begin{equation}
\frac{\delta}{\delta \bar{J} (p)} =
\left(\frac{\Lambda}{\mu}\right)^{-\gamma} \Lambda^{\frac{D+2}{2}}
\frac{\delta}{\delta \JJ (p \Lambda)}\,,
\end{equation}
we obtain
\begin{eqnarray}
&&\int_p \JJ (-p\Lambda) D_\mu^K (p) \Lambda^D \frac{\delta \WW
  [\JJ]}{\delta \JJ (- p \Lambda)} + \int_p \left( - p
  \cdot \partial_p + 2 - 2\gamma \right) R(p)\nn\\
&&\quad \times \frac{1}{2} \int_q \delta (q-p)
\Lambda^{D+2} \left(\frac{\Lambda}{\mu}\right)^{-2\gamma}
\frac{\partial}{\partial p_\mu} \lb
\frac{\delta^2 \WW [\JJ]}{\delta \JJ (p \Lambda) \delta \JJ (-q
  \Lambda)} + \frac{\delta \WW [\JJ]}{\delta \JJ (p \Lambda)}
\frac{\delta \WW [\JJ]}{\delta \JJ (- q\Lambda)} \rb = 0\,.
\end{eqnarray}
Replacing $p\Lambda$ by $p$, and dividing the whole thing by $\Lambda$, we
obtain
\begin{eqnarray}
&&\int_p \JJ (-p) D_\mu^K (p) \frac{\delta \WW [\JJ]}{\delta \JJ
  (-p)}
+ \int_p \Lambda \frac{\partial}{\partial \Lambda} \left(
R_\Lambda (p) \left(\frac{\Lambda}{\mu}\right)^{-2\gamma}\right)\nn\\ 
&&\quad \times \frac{1}{2} \int_q \delta (q-p)
 \frac{\partial}{\partial
  p_\mu}
\lb \frac{\delta^2 \WW [\JJ]}{\delta \JJ (p) \delta \JJ (-q)} +
\frac{\delta \WW [\JJ]}{\delta \JJ (p)} 
\frac{\delta \WW [\JJ]}{\delta \JJ (- q)} \rb = 0\,.
\end{eqnarray}
Since
\begin{equation}
\lim_{\Lambda \to 0+} \Lambda \frac{\partial}{\partial \Lambda} \left(
R_\Lambda (p) \left(\frac{\Lambda}{\mu}\right)^{-2\gamma}\right) =
0\,,
\end{equation}
we obtain (\ref{WW-conformal}) in the limit $\Lambda \to 0+$.

\section{Extension to massive theories\label{section-massive}}

The main result (\ref{mainresult}) can be extended to massive
theories, but the extension is less interesting for the reason we give
at the end of the section.

Let us consider a theory with a mass parameter $g$ with mass
dimension, say, $2$.  In the dimensionful convention, the generating
functional of the connected correlation functions is given by
\begin{equation}
\WW (g) [\JJ] = \lim_{\Lambda \to 0+} W_\Lambda (g) [\JJ] =
\lim_{\Lambda \to 0+} \tilde{W}_\Lambda (g) \left[
  \left(\frac{\Lambda}{\mu}\right)^\gamma \JJ \right]
\end{equation}
from (\ref{WW-Wtilde}).  Note that $g$ is a constant independent of
$\Lambda$, and we have assumed that the anomalous dimension is
independent of $g$.  At $g=0$ we recover the fixed point theory
considered in the previous section.  Let $y > 0$ be the scale
dimension of the mass parameter in the dimensionless convention.  Then 
the dimensionless mass parameter is related to $g$ by
\begin{equation}
\bar{g} \equiv \frac{g}{\mu^2} \left(\frac{\mu}{\Lambda}\right)^y\,.
\label{gbar}
\end{equation}
Since
\begin{equation}
\partial_t \bar{g} = - \Lambda \frac{\partial}{\partial \Lambda}
\bar{g}\Big|_g = y\, \bar{g}\,,
\end{equation}
we can trade $\partial_t$ for $y \bar{g} \frac{\partial}{\partial
  \bar{g}}$. Then, $\bar{W} (\bar{g})$ satisfies
\begin{eqnarray}
y \bar{g} \frac{\partial}{\partial \bar{g}} e^{\bar{W} (\bar{g})[\bar{J}]}
&=& \int_p \left[ \left( p \cdot \partial_p + \frac{D-2}{2} +
  \gamma \right) \bar{J} (p) \cdot
\frac{\delta}{\delta \bar{J} (p)} \right. \nn\\
&&\left. + \left( - p \cdot \partial_p + 2 - 2
  \gamma \right) R(p)  \frac{1}{2} \frac{\delta^2}{\delta
  \bar{J} (p) \delta \bar{J} (-p)} \right] e^{\bar{W}(g) [J]}\,.
\end{eqnarray}
Since
\begin{equation}
\bar{W} (\bar{g}) [\bar{J}] = \tilde{W}_\Lambda (g) [J]\,,
\end{equation}
we obtain, from (\ref{WW-Wtilde}),
\begin{equation}
\WW (g) [\JJ] = \lim_{\Lambda \to 0} \bar{W} (\bar{g})
 \left[ \bar{J} (p) = \left(\frac{\Lambda}{\mu}\right)^\gamma
   \Lambda^{\frac{D-2}{2}} \JJ (p \Lambda) \right]\,,\label{massive}
\end{equation}
where the $\Lambda$-dependence of $\bar{g}$ is given by (\ref{gbar}).
Note that $\bar{g}$ diverges as we take $\Lambda \to 0+$.

For example, consider the simplest example of the massive Gaussian
theory, corresponding to $y=2$.  We obtain
\begin{equation}
\bar{W} \left(\frac{m^2}{\Lambda^2}\right) [\bar{J}]
= \frac{1}{2} \int_p \frac{\bar{J} (p) \bar{J} (-p)}{p^2 +
  m^2/\Lambda^2 + R (p)} \,,
\end{equation}
We then find
\begin{eqnarray}
  \bar{W} \left(\frac{m^2}{\Lambda^2}\right)
  \left[ \Lambda^{\frac{D-2}{2}} \JJ (p \Lambda) \right] 
  &=& W_\Lambda (m^2) [\JJ] 
= \frac{1}{2} \int_p \frac{\JJ (p) \JJ
    (-p)}{p^2 + m^2 + R_\Lambda (p)}\nn\\
&\stackrel{\Lambda \to 0}{\longrightarrow}& \frac{1}{2} \int_p \frac{\JJ
    (p) \JJ(-p)}{p^2+ m^2}\,.
\end{eqnarray}

The crucial difference of (\ref{massive}) from (\ref{mainresult}) is
that the right-hand side is not the high momentum limit of a fixed
$\bar{W}^*$: $\bar{W} (\bar{g})$ depends on the exponentially large
parameter $\bar{g}$.  This is expected.  Take a fixed momentum $p$ for
the left-hand side of (\ref{massive}).  The mass scale is of order
$\mu (g/\mu^2)^{\frac{1}{y}}$.  Now, for the right-hand side, the
corresponding dimensionless momentum is $p/\Lambda$.  Since the ratio
to the mass scale must be the same
\begin{equation}
\frac{p}{\mu (g/\mu^2)^{\frac{1}{y}}} =
\frac{p/\Lambda}{\bar{g}^{\frac{1}{y}}}\,,
\end{equation}
we reproduce (\ref{gbar})
\begin{equation}
\bar{g} = \frac{g}{\mu^2}
\left(\frac{\mu}{\Lambda}\right)^{\frac{1}{y}}
\stackrel{\Lambda \to 0+}{\longrightarrow} + \infty\,.
\end{equation}
To obtain $\bar{W} (\bar{g})$ for large $\bar{g}$, we must solve the
ERG equation for a wide range of $\bar{g}$.  We have nothing to gain
by switching to the dimensionless convention.

\section{Conclusion\label{conclusion}}

In this paper we have shown that the high momentum limit of a fixed
point Wilson action contains the connected correlation functions of
the corresponding massless theory.  This is given explicitly by
(\ref{mainresult}), where $\WW$ is the generating functional of the
connected correlation functions, and $\bar{W}^*$ is the generating
functional with an IR cutoff associated with the fixed point Wilson
action $\bar{S}^*$.  $\bar{W}^*$ is directly related to $\bar{S}^*$ by
\begin{subequations}
\begin{eqnarray}
\bar{W}^* [\bar{J}] &=& \frac{1}{2} \int_p \frac{\bar{J} (-p) \bar{J}
  (p)}{R(p)} + \bar{S}^* [\bar{\phi}] \,,\\
\bar{J} (p) &\equiv& \frac{R (p)}{K(p)} \bar{\phi} (p)\,,
\end{eqnarray}
\end{subequations}
where $K, R$ are cutoff functions.  In deriving (\ref{mainresult}), we
have used two equivalent conventions for ERG: one with dimensionful
cutoff $\Lambda$, and the other with a fixed dimensionless cutoff $1$.
In the dimensionful convention, the correlation functions are obtained
from the Wilson action in the limit of the vanishing cutoff, as given
by (\ref{WW-WL}) and (\ref{WW-Wtilde}).  On the other hand, in the
dimensionless convention, the correlation functions are obtained as
the high momentum limit of the Wilson action.  We have used both
conventions to derive (\ref{mainresult}).

Recently, in \cite{Rosten:2017urs}, a classical limit has been introduced as the
limit of an infinite momentum cutoff where the naive scale and conformal
invariance may be restored in the Wilson action.  We have discussed the
opposite limit of the vanishing cutoff in this paper.

\appendix

\section{Derivation of the diffusion equation\label{appendix-diffusion}}

We have derived a variant of diffusion equation three times from the
corresponding ERG differential equation: (\ref{polchinski-diffusion})
from (\ref{polchinski}), (\ref{diffusion}) from (\ref{ERGdiffeq}), and
(\ref{diffusion-gamma}) from (\ref{ERGdiffeq-gamma}).  The derivation
is essentially the same, and let us show how to derive
(\ref{diffusion-gamma}) from (\ref{ERGdiffeq-gamma}) here.

Differentiating $\tilde{W}_\Lambda [J]$ with respect to $\Lambda$
while fixing $J$, we obtain
\begin{equation}
- \Lambda \frac{\partial}{\partial \Lambda} \tilde{W}_\Lambda [J] = \frac{1}{2}
\int_p \frac{\Lambda \frac{\partial}{\partial \Lambda} R_\Lambda
  (p)}{R_\Lambda (p)^2} J(p) J(-p) - \Lambda \frac{\partial}{\partial
  \Lambda} \tilde{S}_\Lambda [\phi]\Big|_J\,.
\end{equation}
Since
\begin{equation}
\phi (p) = \frac{K_\Lambda (p)}{R_\Lambda (p)} J (p)\,,
\end{equation}
we obtain
\begin{equation}
- \Lambda \frac{\partial}{\partial \Lambda} \tilde{S}_\Lambda [\phi]\Big|_J
= - \Lambda \frac{\partial}{\partial \Lambda} \tilde{S}_\Lambda [\phi] - \int_p
\Lambda \frac{\partial}{\partial \Lambda} \ln \frac{K_\Lambda
  (p)}{R_\Lambda (p)} \cdot \phi (p) \frac{\delta \tilde{S}_\Lambda [\phi]}{\delta
  \phi (p)}\,.
\end{equation}
Using (\ref{ERGdiffeq-gamma}), we obtain
\begin{eqnarray}
&&- \Lambda \frac{\partial}{\partial \Lambda} \tilde{W}_\Lambda [J] =
\frac{1}{2} \int_p \frac{\Lambda \frac{\partial}{\partial \Lambda} R_\Lambda
  (p)}{R_\Lambda (p)^2} J(p) J(-p)  + \int_p
\left( \Lambda \frac{\partial}{\partial \Lambda} \ln R_\Lambda (p) -
  \gamma \right) \phi (p) \frac{\delta \tilde{S}_\Lambda [\phi]}{\delta 
  \phi (p)} \nn\\
&&\quad + \int_p \left( \Lambda \frac{\partial}{\partial \Lambda} R_\Lambda
  (p) - 2 \gamma R_\Lambda (p) \right)
\frac{K_\Lambda (p)^2}{R_\Lambda (p)^2} \frac{1}{2} \lb
\frac{\delta \tilde{S}_\Lambda}{\delta \phi (p)} \frac{\delta
  \tilde{S}_\Lambda}{\delta \phi (-p)} 
+ \frac{\delta^2 \tilde{S}_\Lambda}{\delta \phi (p) \delta \phi
  (-p)}\rb\,. 
\end{eqnarray}
Using
\begin{equation}
\frac{\delta \tilde{S}_\Lambda [\phi]}{\delta \phi (p)} =
\frac{R_\Lambda (p)}{K_\Lambda 
  (p)} \frac{\delta}{\delta J(p)} \left( \tilde{W}_\Lambda [J] - \frac{1}{2}
  \int_p \frac{J (p) J(-p)}{R_\Lambda (p)} \right)\,,
\end{equation}
and ignoring the $J$-independent terms, we obtain
\begin{eqnarray}
&&- \Lambda \frac{\partial}{\partial \Lambda} \tilde{W}_\Lambda [J] = \int_p 
\left[ \gamma J(p) \frac{\delta \tilde{W}_\Lambda [J]}{\delta J(p)}
\right.\nn\\
&&\quad\left. +
  \left( \Lambda 
\frac{\partial R_\Lambda (p)}{\partial \Lambda} - 2 \gamma R_\Lambda
(p) \right) \frac{1}{2} \lb \frac{\delta^2 \tilde{W}_\Lambda
[J]}{\delta J(p) \delta J(-p)} + \frac{\delta \tilde{W}_\Lambda
[J]}{\delta J(p)} \frac{\delta \tilde{W}_\Lambda [J]}{\delta J(-p)} \rb
\right] \,,
\end{eqnarray}
which is (\ref{diffusion-gamma}).

\section{Conversion between the dimensionful and the dimensionless
  conventions\label{appendix-conversion}}

Let us derive the dimensionless diffusion equation
(\ref{diffusion-dimless}) from the dimensionful diffusion equation
(\ref{diffusion-gamma}), where $\bar{W}_t [\bar{J}]$ and
$\tilde{W}_\Lambda [J]$ are related by (\ref{table-conversion}).
Differentiating $\bar{W}_t [\bar{J}]$ with respect to $t$, we are fixing
$\bar{J}$:
\begin{equation}
\partial_t \bar{W}_t [\bar{J}] = - \Lambda \frac{\partial}{\partial
  \Lambda} \tilde{W}_\Lambda [J] \Big|_{\bar{J}}\,.
\end{equation}
Since $J$ and $\bar{J}$ are related by (\ref{JJbar}), we obtain
\begin{equation}
\partial_t \bar{W}_t [\bar{J}] =  - \Lambda \frac{\partial}{\partial
  \Lambda} \tilde{W}_\Lambda [J] + \int_p \left( \frac{D-2}{2} + p
  \cdot \partial_p \right) J(p)\cdot \frac{\delta}{\delta J(p)}
\tilde{W}_\Lambda [J]\,.
\end{equation}
Using (\ref{diffusion-gamma}) and (\ref{ddJ-ddJbar}), we obtain
\begin{eqnarray}
\partial_t e^{\bar{W}_t [\bar{J}]} &=& \int_p 
\left(\frac{D-2}{2} + p \cdot \partial_p + \gamma \right) \bar{J} (p)
\cdot \frac{\delta}{\delta \bar{J} (p)} e^{\bar{W}_t [\bar{J}]}\nn\\
&& + \int_p \left(\Lambda \frac{\partial}{\partial \Lambda} - 2 \gamma
\right) \left(\Lambda^2 R(p/\Lambda)\right) \cdot \frac{1}{2}
\frac{\delta^2}{\delta J(p) \delta J(-p)} e^{\bar{W}_t [\bar{J}]}\nn\\
&=& \int_p 
\left(\frac{D-2}{2} + p \cdot \partial_p + \gamma \right) \bar{J} (p)
\cdot \frac{\delta}{\delta \bar{J} (p)} e^{\bar{W}_t [\bar{J}]}\nn\\
&& + \int_p \Lambda^2 \left(- p \cdot \partial_p + 2 - 2 \gamma
\right) R(p/\Lambda) \cdot \Lambda^{-D-2} \frac{1}{2}
\frac{\delta^2}{\delta \bar{J} (p/\Lambda) \delta \bar{J}
  (-p/\Lambda)}  e^{\bar{W}_t [\bar{J}]} \nn\\
&=& \int_p \left[ \left( p \cdot \partial_p +
\frac{D-2}{2} + \gamma\right) \bar{J} (p)
\cdot \frac{\delta}{\delta \bar{J} (p)} \right.\nn\\
&& \left.\qquad + \left( - p \cdot \partial_p + 2 - 2\gamma\right) R(p)
  \cdot \frac{1}{2} \frac{\delta^2}{\delta \bar{J} (p) \delta \bar{J}
    (-p)} \right]  e^{\bar{W}_t [\bar{J}]}\,,
\end{eqnarray}
which is (\ref{diffusion-dimless}).

\section{Effective action\label{appendix-Gamma}}

The effective action is defined as the Legendre transform of the
generating functional of connected correlation functions:
\begin{subequations}
\label{Gamma-def}
\begin{equation}
\Gamma_{\mathrm{eff}} [\Phi] \equiv \WW [\JJ] - \int_p \JJ (-p) \Phi (p)\,,
\end{equation}
where
\begin{equation}
\Phi (p) \equiv \frac{\delta \WW [\JJ]}{\delta \JJ (-p)}\,.
\end{equation}
\end{subequations}
On the other hand, the so called effective average action
$\bar{\Gamma}$ is defined as the analogous Legendre transform:
\begin{subequations}
\begin{equation}
\bar{\Gamma} [\bar{\Phi}] - \frac{1}{2} \int_p R (p) \bar{\Phi} (-p)
\bar{\Phi} (p) \equiv \bar{W} [\bar{J}] - \int_p \bar{J} (-p)
\bar{\Phi} (p)\,,
\end{equation}
where
\begin{equation}
\bar{\Phi} (p) \equiv \frac{\delta \bar{W}[\bar{J}]}{\delta \bar{J}
  (-p)}\,.
\end{equation}
\end{subequations}
We have omitted the ${}^*$ from $\bar{\Gamma}$ and $\bar{W}$ to
simplify the expression.  We wish to express $\Gamma_{\mathrm{eff}}$
as the IR limit of $\bar{\Gamma}$ by rewriting the main result
(\ref{mainresult}).

Recall Eq.~(\ref{mainresult}) is the IR limit of
\begin{equation}
\WW [\JJ] = \bar{W} [ \bar{J} ]\,,\label{mainresult-nonzeroL}
\end{equation}
where
\begin{equation}
\bar{J} (p) = \left(\frac{\Lambda}{\mu}\right)^\gamma
\Lambda^{\frac{D-2}{2}} \JJ (p \Lambda)\,.
\end{equation}
Correcting (\ref{ddJ-ddJbar}) by the anomalous dimension, we obtain
\begin{equation}
\frac{\delta}{\delta \JJ (-p)} =
\left(\frac{\Lambda}{\mu}\right)^\gamma \Lambda^{-\frac{D+2}{2}}
\frac{\delta}{\delta \bar{J} (-p/\Lambda)}\,.
\end{equation}
Hence, we obtain
\begin{equation}
\Phi (p) = \frac{\delta \WW [\JJ]}{\delta \JJ (-p)} 
= \left(\frac{\Lambda}{\mu}\right)^\gamma \Lambda^{-\frac{D+2}{2}}
\frac{\delta \bar{W}[\bar{J}]}{\delta \bar{J} (-p/\Lambda)}
= \left(\frac{\Lambda}{\mu}\right)^\gamma \Lambda^{-\frac{D+2}{2}}
\bar{\Phi} (p/\Lambda)\,.
\end{equation}
Thus, from (\ref{Gamma-def}), we obtain
\begin{eqnarray}
\Gamma_{\mathrm{eff}} [\Phi] &=& \bar{W} [\bar{J}] - \int_p \JJ (-p)
\Phi (p)\nn\\ &=& \bar{W} [\bar{J}] - \int_p
\left(\frac{\Lambda}{\mu}\right)^{-\gamma} \Lambda^{-\frac{D-2}{2}}
\bar{J} (-p/\Lambda) \left(\frac{\Lambda}{\mu}\right)^{\gamma}
\Lambda^{-\frac{D+2}{2}} \bar{\Phi} (p/\Lambda) \nn\\
&=& \bar{W} [\bar{J}] - \int_p \bar{J} (-p) \bar{\Phi} (p)\nn\\
&=& \bar{\Gamma}[\bar{\Phi}] - \frac{1}{2} \int_p R(p) \bar{\Phi} (-p)
\bar{\Phi} (p)\,.
\end{eqnarray}
Since
\begin{equation}
\int_p R(p) \bar{\Phi} (-p) \bar{\Phi} (p) = \int_p
\left(\frac{\Lambda}{\mu}\right)^{-2\gamma} \Lambda^2 R(p/\Lambda)
\Phi (-p) \Phi (p)
\end{equation}
vanishes in the limit $\Lambda \to 0+$ as a functional of $\Phi$, we
obtain
\begin{subequations}
\begin{equation}
\Gamma_{\mathrm{eff}} [\Phi] = \lim_{\Lambda \to 0+} \bar{\Gamma} [\bar{\Phi}]\,,
\end{equation}
where
\begin{equation}
\bar{\Phi} (p) = \left(\frac{\Lambda}{\mu}\right)^{-\gamma}
\Lambda^{\frac{D+2}{2}} \Phi (p \Lambda)\,.
\end{equation}
\end{subequations}
This is the desired result.



\bibliography{version2}

\end{document}